
\def\d#1/d#2{ {\partial #1\over\partial #2} }
\newcount\sectno

\def\sect{\global\advance\sectno by1 \the\sectno }
\def\pdr{\partial}

\def\be{\beta}
\def\ga{\gamma}

\def\eps{\epsilon}
\def\half{{1\over 2}}
\def\tr{\hbox{tr}}

\newcount\eqnumber
\def\beq{ \global\advance\eqnumber by 1 $$ }
\def\eeq{ \eqno(\the\eqnumber)$$ }
\def\n{\global\advance \eqnumber by 1\eqno(\the\eqnumber)}
\def\puteqno{\global\advance \eqnumber by 1 (\the\eqnumber)}
\def\beqs{$$\eqalign}
\def\eeqs{$$}
\newcount\refno\refno=0  
\def\ifundefined#1{\expandafter\ifx\csname #1\endcsname\relax}

\def\[#1]{
\ifundefined{#1}\advance\refno by
1\expandafter\edef\csname#1\endcsname{\the\refno}\fi[\csname #1\endcsname]}

\def\refis#1{\noindent\csname #1\endcsname. }

\def\label#1{
\ifundefined{#1}
\expandafter\edef\csname #1\endcsname{\the\eqnumber}
\else\message{label #1 already in use}
\fi{}}
\def\(#1){(\csname #1\endcsname)}
\def\eqn#1{(\csname #1\endcsname)}

\baselineskip=15pt
\parskip=10pt

\magnification=1200


\def\tpsi{ {\tilde \psi} }
\def\talpha{ {\tilde \alpha} }

\rightline{ UR-1298 \ \ \ \ \ \ \ \ \ }
\rightline{ ER-40685-747  \ \ }
\rightline{ hep-ph 9301208 }
\rightline{ March 1993   \ \ \ \ \ }

\vskip.4in
\centerline{\bf Spherical Quantum Chromodynamics of Heavy Quark Systems}
\vskip.4in
\centerline{\rm   K. S. Gupta, S. Guruswamy and  S. G. Rajeev}
\vskip.4in
\centerline{\it Department of Physics and Astronomy}
\centerline{\it University of Rochester}
\centerline{\it Rochester, N.Y. 14627}
\centerline{\it e-mail: gupta@uorhep, guruswamy@uorhep and rajeev@uorhep}
\vskip.9in

\centerline{\bf Abstract}

We propose a model for Quantum Chromodynamics, obtained by ignoring the
angular
 dependence of the gluon fields, which could qualitatively describe systems
 containing one heavy quark. This leads to a two
 dimensional  gauge theory which has chiral symmetry and  heavy quark symmetry.
 We show that in a light cone formalism, the Hamiltonian of this spherical QCD
 can be
 expressed entirely in terms of color singlet variables. Furthermore, in the
 large $N_c$ limit, it tends to a   classical hadron theory. We derive an
 integral equation for the masses and wavefunctions of a heavy meson. This can
 be  interpreted as a relativistic potential model.
The integral  equation is scale invariant, but  renormalization of the coupling
 constant generates a scale. We compute the approximate beta function of the
 coupling constant, which has an ultraviolet stable fixed point at the origin.

\vfill\eject

{\it \sect. Introduction}
\def\slash#1{ {\gamma\cdot #1}}

Two dimensional Quantum Chromodynamics (QCD)  is fairly well
understood in
the ${1\over N_c}$ expansion, following the work of 't Hooft, Witten and
others \[thooft],\[witten],\[N]. Furthermore, it can  be solved by
numerically diagonalizing the hamiltonian\[pauli]. The two methods are in
satisfactory agreement with each other.
 In this paper we will study an
approximation to four dimensional QCD in which the gluon field is independent
 of two angular  variables, so that two dimensional  methods can be applied.
Although this could be  studied just as a toy model for QCD, it might also be a
model for hadrons containing one heavy quark. These systems have attracted
much attention recently with the discovery of a new heavy quark
symmetry\[iw],\[georgi],\[falk].
Such a hadron is similar to an
atom, the heavy quark being like the nucleus with the light quarks orbiting
like electrons around it. The main difference is that the
light quarks have to be described relativistically. Still, we should be able to
treat these systems in the spirit of the Hartree--Fock or Thomas--Fermi
methods\[slater]. The gluon field produced by the heavy quark is spherically
symmetric. That of the light quarks is not spherically symmetric because the
light quark wavefunctions depend on angles.  Still, we should be able to do a
`spherical
 averaging' as in Hartree--Fock theory and approximate the current density of
 the
light quarks by their average over the angular variables.  The average of the
angular components of the current density will vanish.  This spherical
averaging is equivalent to putting $A_{\theta}=A_{\phi}=0$ and assuming that
the remaining components $A$ are independent of the angles. This
will
lead to an effectively two dimensional theory ( although without translation
invariance) in light cone coordinates. This `Spherical QCD'
respects heavy quark symmetry, chiral symmetry of light quarks and scale
invariance. (The scale invariance should be broken by  quantum
effects.) It should be possible to solve this theory numerically, using methods
similar to those used for two dimensional QCD\[pauli]. This is vastly simpler
than solving the full $3+1$ dimensional problem numerically.
 In this paper, we will study this spherical QCD  analytically using the
${1\over N_c}$
expansion.

In this paper, we will view Spherical QCD as a model to study theoretical
 aspects of QCD such as the ${1\over N_c}$ expansion and asymptotic freedom.
We have a model that is solvable in the large $N_c$ limit, in which the beta
 function can be calculated in a non--perturbative approximation. Whether our
 approximate theory  can make reliable numerical predictions can only be seen
 from future work.

't Hooft's original approach  to the ${1\over N_c}$  expansion involved summing
planar Feynman diagrams. This method is difficult to generalize to our
 situation. More
recently, another method has been used to construct  a bilocal quantum field
 theory of hadrons equivalent to 2DQCD for
all $N_c$ \[twodmeson]. The parameter ${1\over
N_c}$ plays the role of $\hbar$ in this hadron theory ( $N_c$ must be an
 integer
for topological reasons) so that the large $N_c$ limit of QCD corresponds to a
classical theory of hadrons. The field equations of this theory are certain
{\it  non-linear} integral equations: the large $N_c$ limit of
QCD is not a free field theory.  Semiclassical expansion around the classical
solutions will describe the hadron masses and wavefunctions in the ${1\over
N_c}$ expansion. The most obvious  static solution is the vacuum, and small
fluctuations around it (mesons) are described by a linear integral equation, in
agreement with 't Hooft. This point of view also allows for baryons, which are
then static solutions that deviate by a large amount from the
vacuum\[twodmeson].

This approach to two dimensional QCD can be generalized to Spherical QCD.
One of our results is a linear integral equation for the heavy--light meson
 masses and
wave functions. We can understand our result ( just like  't Hooft's integral
equation) as a relativistic potential model. Ordinarily, such a one particle
point of view would not be allowed in this highly relativistic situation  due
to
 effects of virtual light $q\bar q$ pairs. In the meson picture, these are the
effects  of virtual light mesons.
However, in the large $N_c$ limit such effects are suppressed, so that  a one
particle description  is allowed. Thus we will get  a light quark moving in the
 field of the heavy quark. The Dirac equation can be separated in
light cone coordinates, reducing the problem to a linear integral equation.
(That light cone coordinates are better suited to nonperturbative problems in
QCD have also been suggested  for other reasons
\[brodsky],\[wilson]. ) It should be possible to  fit meson masses to a
potential with a
small number of parameters as a phenomenological test of our approach.

In the next section (Sec.2 ) we will derive the action of spherical QCD; it
will
 be convenient to use a light cone coordinate system. In Sec. 3 we will show
 that this theory can be written entirely in terms of a set of color singlet
 fields
representing hadrons. The commutation relations and equations of motion of
these
 variables are derived. In Sec. 4 we show that in the large $N_c$ limit, this
 tends to a classical theory whose Poisson brackets, constraints and equations
 of motion are derived.  By expanding around the vacuum solution we get a
linear
 integral  equation for the masses and wavefunction of heavy mesons.
In Sec 5. we have analysed the  ground state of the wave equation in a
 variational approach. A renormalization of the coupling constant is necessary
 to keep the ground state energy finite as the cut--off is removed. We compute
 the approximate beta function and show that it has an ultraviolet stable fixed
 point at the origin. Finally,
in
 the appendix we describe how the constraints on the color singlet variables
can
 be derived, in a simplified context.

\vfill\eject
{\it \sect. QCD in light cone coordinates}\hfill\break

The action of four dimensional QCD  with $N_c$ colors, $n_f$ massless quarks
and $N_f$
heavy quarks of mass $M$ is

\beqs{
        S\;\;& =-{1\over 2 g^2}\int \tr F_{\mu\nu}F^{\mu\nu} d^4x
+\int \bar\psi^a[i\ga\cdot\pdr+ \ga\cdot A]\psi_a d^4x\cr
\;\;&+\int \bar\Psi^A[i\ga\cdot\pdr+ \ga\cdot A+M]\Psi_A d^4x,\cr
}\eeqs
where $A_\mu$ and $F_{\mu \nu}$ are $N_c$ x $N_c$ hermitian matrices.
Here, the color indices are suppressed and $a=1,\cdots n_f$, $A=1,\cdots N_f$.
We are interested in systems that contain only one heavy quark, which can be
assumed to be at rest at the origin.
Conventional spherical polar coordinates $t,r,\theta,\phi$ are cumbersome in
this context due to the negative energy states of the light fermions and the
dependence of the Dirac sea on the gluon fields. We find it most convenient to
introduce a sort of light cone coordinate system  $u,r,\theta,\phi$ centered
 at the
heavy quark:
\beq
        u=t+r,\quad ds^2=du(du-2dr)-r^2(d\theta^2+\sin^2\theta d\phi^2).
\eeq
Note that this is not an orthogonal coordinate system; yet the Dirac equation
is separable in this system. The surface $u=$constant is a past light cone with
apex at the origin. We will use a canonical formalism in which  initial data is
given on this null surface. The vector field ${\pdr\over \pdr u}$ is in fact
just
the usual time translation, so that its conjugate variable  is energy.
But ${\pdr\over \pdr r}$ for fixed $u$ is {\it  not} the same as the radial
 vector in
spherical polar coordinates. Instead, it is a {\it  null}
vector pointing outward along the surface $u=$constant:
\beq
        \left({\pdr\over \pdr r}\right)_{t}=\left({\pdr\over \pdr
r}\right)_{u}+{\pdr\over \pdr u}.
\eeq

We will expand all the fields in appropriate spherical harmonics in the angular
variables. The essential approximation we will make is to  include only the
s-wave states of the gluon fields.
 This amounts to replacing the current density of the light quarks by its
average over the angular variables, so that the mean gluon field they produce
is spherically symmetric.( This concept of a `spherical average' was introduced
into atomic physics by Hartree--Fock and is extensively used in calculations of
atomic energy levels and wavefunctions\[slater].) We will be
able to eliminate the gluon fields and half the degrees of freedom of the light
quarks, to get a theory of quarks interacting through a selfconsistent
potential. The resulting nonlocal lagrangian will have chiral symmetry, heavy
quark symmetry and scale invariance, and defines our `Spherical QCD'.

Let us first simplify the gluon Lagrangian. After  spherical averaging,
the angular components of a spin one field must vanish,
and $A_u$ and $A_r$ would be independent of the angles. Furthermore we
can choose the gauge condition
\beq
        A_r=0,
\eeq
leaving only the component $A_u=A$.
In this case the Yang--Mills field strength is
\beq
        F_{\mu\nu}dx^\mu \wedge dx^{\nu}=\pdr_r A_u dr\wedge du
\eeq
and the Yang--Mills action is
\beq
        S_{\rm YM}= {1\over \alpha}\int\tr[\pdr_r A]^2r^2drdu.
\eeq
( Here $\alpha={g^2\over 4\pi}$.)
In this gauge there are no propagating components for the Yang--Mills fields
since there is no derivative of $A$ with respect to $u$ in the action.

Let us now turn to the light quark action.
The free Dirac operator in the usual spherical polar coordinates is \[dirac]
\beq
        i\ga\cdot\pdr=i\ga_0[\d/dt+\alpha_r\{\d/dr+{1\over r}-{K\over r}\}]
\eeq
where
\beq
        K={\bf \Sigma}\cdot {\bf L}+1,\quad \alpha_r^2=1\quad [\alpha_r,K]_+=0.
\eeq
As usual ${\bf \Sigma}$ are the Dirac spin matrices, $\alpha_r=\ga_0\gamma_r$,
and ${\bf L}$ is the orbital angular momentum operator. In our coordinates,
\beq
        i\ga\cdot\pdr=i\ga_0[(1+\alpha_r)\d/du+\alpha_r\{\d/dr+{1\over
r}-{K\over r}\}].
\eeq
 It will be convenient to diagonalize $\alpha_r$ and  $\ga_5$ (we define
$\ga_5$ such that  $\ga_5^2=1$):  \beq
        \ga_5 v_{k,\eps,\mu}=\eps v_{k,\eps,\mu}\quad
\alpha_r v_{k,\eps,\mu}=\mu v_{k,\eps,\mu}\quad
Kv_{k,\eps,\mu}=kv_{k,\eps,-\mu}.
\eeq
The eigenvalue $k$ of $K$ is a positive integer.The spinorial
harmonics $v_{k,\eps,\mu}$ are orthogonal and are normalized such that
\beq
\int d\Omega |v_{k,\eps,\mu}|^2=4\pi.
\eeq
Then $\psi$ can be expanded in this basis (suppressing color and flavor indices
on $\psi$ and $\chi$):
\beq
        \psi=\sum_{\mu=\pm1,k>0,\eps=\pm1}
{1\over \surd (8\pi)\;r}\chi_{k,\eps,\mu}(r)v_{k,\eps,\mu}.
\eeq
The factor ${1\over \surd (8\pi)\; r}$ is chosen so as to simplify later
expressions.
In this basis, $\alpha_r=\sigma_3$,$K=k\sigma_1$ and $-i\alpha_rK=k\sigma_2$.
The free Dirac action is then
\beq
\int \bar\psi[i\ga\cdot\pdr]\psi d^4x=   {1\over 2}\sum_{k,\eps}\int
drdu\chi^{\dag}_{k\eps}[i(1+\sigma_3)\d/du+i\sigma_3\d/dr+{k\over
r}\sigma_2]\chi_{k\eps},
\eeq
where $\chi_{k,\eps}=\pmatrix{\chi_{k,\eps,1}\cr\chi_{k,\eps,-1}}$.
Including the gauge field amounts to replacing $\pdr_u$ by $\pdr_u-iA$:
\beq
\int \bar\psi[i\ga\cdot\pdr+\ga\cdot A]\psi d^4x=   {1\over
2}\sum_{k,\eps}\int
drdu\chi^{\dag}_{k\eps}[i(1+\sigma_3)[\d/du-iA]+i\sigma_3\d/dr+{k\over
r}\sigma_2]\chi_{k\eps}.
\eeq

The heavy quark action can be simplified by the transformation\[iw],\[georgi]
\beq
        \Psi\to e^{iMv\cdot x} \Psi,
\eeq
so that in the limit $M\to \infty$
\beq
        \int \bar \Psi^A[i\ga\cdot\pdr+ \ga\cdot A+M]\Psi_A d^4x\to
i\int \bar \Psi^A v\cdot [\pdr-iA]\Psi_A d^4x.
\eeq
Here $v$ is the heavy quark velocity. Now we put
\beq
        \Psi_A={1\over \surd(4\pi) r} Q_A;
\eeq
 the spinor $Q$ satisfies
${\slash v }Q=Q$ \[iw],\[georgi]
and depends only on $r$. In our coordinate  system,
 $v=(1,0,0,0)$ so that this
 becomes just
\beq
i \int  Q^{\dag}[\pdr_u-iA] Q drdu.
\eeq
(We may suppress flavor indices as well when they are not essential).

Thus the complete action is
\beqs{
        S\;\;& =i \int
Q^{\dag}[\pdr_u-iA] Q drdu+{1\over \alpha}\int\tr[\pdr_r
 A]^2r^2drdu\cr
\;\;& +{1\over 2}\sum_{k,\eps}\int
drdu\chi^{\dag}_{k\eps}[i(1+\sigma_3)\{\d/du-iA\}+i\sigma_3\d/dr+{k\over
r}\sigma_2]\chi_{k\eps}.\cr
}\eeqs

In this action, $A$ and $\chi_{k,\eps,-1}$ do not have  derivatives with
respect to $u$. Hence they do not propagate and can be eliminated by solving
their equations of motion. We have,
\beq
        \pdr_r[r^2\pdr_r
A_{\alpha}]=\alpha\rho_{\alpha}
\eeq
where,
\beq
\rho_{\alpha}=\sum_{k,\eps}\chi_{k,\eps,1}^{\dag}t_{\alpha}\chi_{k,\eps,1}
 +Q^{\dag} t_{\alpha}Q,
\eeq
and  $t_{\alpha}$ are the color matrices.  We normalize them such that
\beq
        \tr t_{\alpha}t_{\beta}={1 \over 2}\delta_{\alpha\beta}.
\eeq
Half the light quark degrees of freedom also do not propagate:
\beq
-i\d/dr\chi_{k,\eps,-1}+i{k\over r}\chi_{k,\eps,1}=0.
\eeq
Then,
\beq
        A(r)=\alpha\int_0^{\infty} G(r,r'){\rho(r')}dr'
\eeq
where
\beq
        G(r,r')=-{\rm min} ({1\over r},{1\over r'})
\eeq
and
\beq
        \chi_{k,\eps,-1}=k\pdr^{-1}{\chi_{k\eps,1}\over r}=\int_{\infty}^{r}
{k\over r'}\chi_{k,\eps,1}(r')dr' .
\eeq
The boundary condition are that as $r\to \infty$  the fields vanish. That is,
there is no incoming radiation from past  null infinity.

We can now eliminate $A,\chi_{k\eps,-1}$ to get an action depending only on
 $\chi_{k\eps 1}$ and $Q$:
\beqs{
        S\;\;& =i\int Q^{\dag}\dot Qdrdu+
i\sum_{k\eps}\int drdu\chi^{\dag}_{k\eps 1}\dot
 \chi_{k\eps 1}\cr
\;\;& + \half \sum_{k\eps}\int drdu\chi^{\dag}_{k\eps 1}[
i\pdr_r+k^2r^{-1}(i\pdr_r)^{-1}r^{-1}]\chi_{k\eps 1} - \half\alpha\int
 dr'drdu\rho_{\alpha}(r)\rho_{\alpha}(r')G(r,r')\cr
\;\;& +\alpha\int du dr dr' G(r,r')Q^{\dag}(r)t_{\alpha}\rho_{\alpha}(r')Q(r)
+\alpha  \int du dr dr' G(r,r')\chi^{\dag}_{k\eps 1}(r)
t_{\alpha}\rho_{\alpha}(r')\chi_{k\eps 1}(r)
}\eeqs
This action defines Spherical QCD. It is manifestly invariant under
 $U(n_f)_L\times U(n_f)_R$ ( chiral symmetry), $SU(2N-f)$ ( spin and flavor
 symmetry of heavy quarks) and scale transformations ( the only coupling
 constant $g^2$ is dimensionless). The first two terms simply say that
 $Q,\chi_{k\eps 1}$ are canonically conjugate to $\bar Q,\chi^{\dag}_{k\eps
1}$.
\beq
        [Q^{\dag}(r),Q(r')]_+=\delta(r-r'),\quad
  [\chi_{k\eps
1}^{\dag}(r),\chi_{k'\eps'1}(r')]_+=\delta_{kk'}\delta_{\eps\eps'}\delta(r-r').
\eeq

 The remaining terms determine the hamiltonian:
\beqs{
 H\;& =\sum_{k\eps}\int dr\chi^{\dag}_{k\eps 1}h\chi_{k\eps 1}+
 \half \alpha\int drdr'\rho_{\alpha}(r)\rho_{\alpha}(r')G(r,r')\cr
\;\;& - \alpha\int drdr'G(r,r')Q^{\dag}(r)t_{\alpha}\rho_{\alpha}(r')Q(r)
- \alpha \int drdr'G(r,r')
\chi^{\dag}_{k\eps 1}(r)t_{\alpha}\rho_{\alpha}(r')\chi_{k\eps 1}(r),
}\eeqs
where
$$h=- \half (i\pdr_r+k^2r^{-1}(i\pdr_r)^{-1}r^{-1})$$
is the single particle hamiltonian of the light quarks. Since
the hamiltonian $H$ is conjugate
to $u$, it is just the energy.
 The above anticommutation relations along with the hamiltonian define
Spherical QCD.

Apart from the fact that $h$ and $G$ have different explicit forms ( and the
 presence of the heavy quarks), this is exactly like the hamiltonian of 2DQCD,
 after eliminating the gluons and half the quark fields. Thus from this
point,we
 can follow the method described in reference \[twodmeson] to turn this into a
 theory of bilocal color singlet ( meson) fields.

{\it \sect. Hadron Theory}

Let us define the  color singlet bilinears,
\beqs{
        H^I_i(r',r)={1\over N_c}Q^{\dag I}(r')\chi_{i}(r)\;\;&,\quad
H^{i\dag}_I(r',r)={1\over N_c}\bar\chi^{i\dag}(r') Q_I(r),
\cr
\quad P^I_J(r,r')={1\over N_c}Q^{\dag I}(r) Q_J(r')\;\;&,\quad
M^i_j(r,r')={1\over
N_c}:\chi^{i\dag}(r)\chi_j(r'):\cr
}\eeqs
where $i=(k,\eps,a)$ ranges over spin,chirality and flavor of the light quark
 while $I$ labels the
spin and flavor of the heavy quark. $P$ is normal ordered with respect to the
 trivial vacuum; after the transformations we made the heavy quark has no
 negative energy states.
The normal ordering of $M$ is defined with
respect to the vacuum obtained by filling the negative energy states of the
one--particle hamiltonian $h=-\half[i\pdr_r+k^2  r^{-1}(i\pdr_r)^{-1}r^{-1}]$.
In fact
\beqs{
{1\over N_c}:\chi^{i\dag}(r)\chi_j(r'):=\;\;&\;\;{1\over
 N_c}\chi^{i\dag}(r)\chi_j(r')
+\half[\eps(r,r') - \delta(r,r')]\delta^i_j \cr
=\;\;&\;\;{1\over 2N_c}[\chi^{i\dag}(r),\chi_j(r')]
+\half\eps(r,r')\delta^i_j.\cr
}\eeqs
 Here $\eps(r,r')$ is the operator that  is $+1$ on the eigenstates of $h$ of
positive energy and $-1$ on those of negative energy:
\beq
\eps(r,r')=\int_0^{\infty}d\lambda u_\lambda(r)u^*_\lambda(r')-
\int^0_{-\infty}d\lambda u_\lambda(r)u_\lambda^*(r').
\eeq
The eigenstates $u_{\lambda}$of $h$ are defined  as below:
\beqs{
-\half [i\pdr_r+k^2 r^{-1}(i\pdr_r)^{-1}r^{-1}]u_{\lambda}(r)\;\;&=\lambda
u_{\lambda}(r)\cr
\int_{-\infty}^{\infty}u_{\lambda}(r)u_{\lambda}^*(r')d\lambda\;\;&=
\delta(r-r').\cr
}\eeqs
Also,  we will often use the projection operators to the positive and negative
 energy states
\beq
        \delta_+(r,r')=\half [\delta(r-r')+\eps(r,r')]=
        \int_0^{\infty}u_{\lambda}(r)u^*_{\lambda}(r') d\lambda
\eeq
\beq
        \delta_-(r,r')=\half [\delta(r-r')-\eps(r,r')]=
        \int_{-\infty}^0 u_{\lambda}(r)u^*_{\lambda}(r') d\lambda.
\eeq

 We will rewrite the commutation relations
and the hamiltonian entirely in terms of these color singlet  variables.
The commutation relations are
\beqs{
[H^I_i(r,r'),H^{j\dag}_J(s,s')] \;\;& =\; {1\over N_c}[
P^I_J(r,s')\delta^j_i\delta(r'-s)-M^j_i(s,r')\delta^I_J\delta(s'-r)\cr
 \;\;&-  \delta_{-}(s,r')\delta^j_i\delta^{I}_J\delta (s'-r)],\cr
[H^{I}_{i}(r,r'),M^{j}_k(s,s')]
\;\;& = {1\over N_c}
H^I_k(r,s')\delta^j_i\delta (r'-s),\cr
[H^I_i(r,r'),P^{J}_K(s,s')] \;\;& = -{1\over N_c}
H^J_i(s,r')\delta^I_K\delta(r-s'),\cr
[H^{i\dag}_I(r,r'),P^{J}_K(s,s')] \;\;& = {1\over N_c}
H^{i\dag}_K(r,s')\delta^J_I\delta (r'-s),\cr
[H^{i\dag}_I(r,r'),M^{j}_k(s,s')] \;\;& = -{1\over N_c}
H^{j\dag}_I(s,r')\delta^i_k\delta (r-s'),\cr
[P^J_I(r,r'),P^K_L(s,s')] \;\;& = {1\over N_c}[ P^J_L(r,s')\delta^K_I
\delta (r'-s) - P^K_I(s,r')\delta^J_L\delta (s'-r)],\cr
[M^j_i(r,r'),M^k_l(s,s')] \;\;& = {1\over N_c}[M^j_l(r,s')\delta^k_i\delta
(s-r')
- M^k_i(s,r')\delta^j_l\delta (r-s'),\cr
 \;\;& +\half(-\eps(r,s')\delta (s-r') + \eps(s,r')\delta
(r-s'))\delta^j_l\delta^k_i].\cr
}\eeqs
The c-number terms proportional to $\eps(r,r')$  and $\delta_{-}$ arises
because
 of the normal
ordering of $M$  and $P$.

The hamiltonian can be expressed  in terms of these bilinears after a Fierz
reordering using the identity for the color matrices:
\beq
        t_{\alpha q}^p t_{\alpha s}^r= {1 \over 2}\delta^p_s\delta^r_q.
\eeq
(We are
using the gauge group $U(N_c)$ rather than $SU(N_c)$, as is usual in large
$N_c$ approaches\[thooft].)
We get, with ${\tilde \alpha}={\alpha} N_c$
\beqs{
{H\over N_c}\;\;&= \int h(r,r')M^i_i(r,r') drdr'+
{1 \over 4} {\tilde \alpha}\int G(r,r')[P^I_J(r,r')P^J_I(r',r)\cr
\;\;& +H^I_i(r,r')H^{i\dag}_I(r',r)+H^{i\dag}_I(r,r')H^I_{i}(r',r)
+M^i_j(r,r')M^j_i(r',r)]drdr'.\cr
}\eeqs

It should be possible to find the eigenvalues of this hamiltonian by
numerically diagonalizing a finite dimensional approximation to it. This is the
method used
in Ref. \[pauli] to solve two dimensional QCD.

Now we note that $N_c$ appears only as an overall constant in the commutation
relations and the hamiltonian. In particular, the equations of motion are
independent of $N_c$.  This means that the limit $N_c\to \infty$ is a
classical limit in which the commutators get replaced by Poisson brackets.

 The large $N_c$ limit of spherical  QCD is a classical hadron theory. The
semiclassical expansion around a solution of this classical theory will be
equivalent to the ${1\over N_c}$ expansion of spherical QCD.

{\it \sect. Classical Hadron Theory}

The Poisson brackets obtained by the correspondence principle
\beq
        [A,B]={i\over N_c}\{A,B\}
\eeq
where ${1\over N_c}$ plays the role of $\hbar$. Thus,
\beqs{
i\{H^I_i(r,r'),H^{j\dag}_J(s,s')\} \;\;& =\;
P^I_J(r,s')\delta^j_i\delta(r'-s)-M^j_i(s,r')\delta^I_J\delta(s'-r)\cr
 \;\;&- \delta_{-}(s,r')\delta^j_i\delta^{I}_J\delta (s'-r),\cr
i\{H^{I}_{i}(r,r'),M^{j}_k(s,s')\}
\;\;& =
H^I_k(r,s')\delta^j_i\delta (r'-s),\cr
i\{H^I_i(r,r'),P^{J}_K(s,s')\} \;\;& = -
H^J_i(s,r')\delta^I_K\delta(r-s'),\cr
i\{H^{i\dag}_I(r,r'),P^{J}_K(s,s')\} \;\;& =
H^{i\dag}_K(r,s')\delta^J_I\delta (r'-s),\cr
i\{H^{i\dag}_I(r,r'),M^{j}_k(s,s')\} \;\;& = -
H^{j\dag}_I(s,r')\delta^i_k\delta (r-s'),\cr
i\{P^J_I(r,r'),P^K_L(s,s')\} \;\;& =  P^J_L(r,s')\delta^K_I
\delta (r'-s) - P^K_I(s,r')\delta^J_L\delta (s'-r),\cr
i\{M^j_i(r,r'),M^k_l(s,s')\} \;\;& = M^j_l(r,s')\delta^k_i\delta
(s-r')
- M^k_i(s,r')\delta^j_l\delta (r-s'),\cr
 \;\;& +\half (-\eps(r,s')\delta (s-r') + \eps(s,r')\delta
(r-s'))\delta^j_l\delta^k_i,\cr
}\eeqs
and the hamiltonian
\beqs{
H &= \int h(r,r')M^i_i(r,r') drdr'+
\half {\tilde \alpha}\int G(r,r')[P^I_J(r,r')P^J_I(r',r)\cr
 & + H^I_i(r,r')H^{i\dag}_I(r',r)
+ H^{i\dag}_I(r,r')H^I_{i}(r',r)
+ M^i_j(r,r')M^j_i(r',r)]drdr'\cr
}\eeqs
define a classical theory.
The equations motion of this classical theory can be worked out by a straight
forward ( but tedious) calculation of the Poisson brackets of the observables
with the hamiltonian. We get
\beqs{
i\d H^I_i(r,r')/d u
=& \half{\tilde\alpha}\int ds' H^I_i(r,s')G(s',r)\delta_{-}(s',r') + \int  ds'
 H^I_i(r,s')h(s',r')\cr
 & - \half \tilde \alpha \int ds' G(r',s')[M^j_i(s',r')H^I_j(r,s')
 + P^I_J(r,s')H^J_i(s',r')]\cr
 & + \half \tilde \alpha \int ds' G(s',r)[P^I_J(r,s')H^J_i(s',r')
+ M^j_i(s',r')H^I_j(r,s')],\cr
i \d {M^i_j(r,r')}/d{u}
&=  {1 \over 4}\tilde \alpha
\int ds' [ G(s',r')M^i_j(s',r')\eps (r,s') - G(r,s')M^i_j(r,s')
    \eps (s',r')]\cr
 & - \half \tilde \alpha \int ds' G(s',r')
[ M^l_j(s',r')M^i_l(r,s')
 + H^I_j(s',r')H^{\dag i}_I(r,s')] \cr
& + \half \tilde \alpha \int ds' G(r,s')
[ M^l_j(s',r')M^i_l(r,s')
+ H^I_j(s',r')H^{\dag i}_I(r,s')]\cr
& + \int ds [  M^i_j(r,s) h(s,r')-h(r,s)M^i_j(s,r')],\cr
i\d P^I_J(r,r')/d u
=& \half \tilde \alpha \int ds' G(r,s')[P^I_K(r,s')P^K_J(s',r') +
 H^I_i(r,s')H^{\dag i}_J(s',r')]\cr
& -\half \tilde \alpha \int ds' G(s',r')
[P^I_K(r,s')P^K_J(s',r')
 + H^I_i(r,s')H^{\dag i}_J(s',r')].\cr
}\eeqs

These equations of motion however do not describe the classical hadron theory
completely. The bilinears satisfy some constraints, whose origin is described
in the appendix in a simplified fermion theory. The constraint for $M$ has been
derived in the appendix. It becomes in our present notation,
\beqs{
        \int M^i_j(r,r')M^j_k(r',r'') dr'-\;\;&\;\; \half\int
[\eps(r,r')M^i_k(r',r'')+ M^i_k(r,r')\eps(r',r'')] dr'\cr
&+\int H^{\dag i}_I(r,r')H^{I}_k(r',r'')dr'=0.\cr
}\eeqs
Similar arguments show that
\beqs{
   \int H^I_i(r,r')M^i_j(r',r'')dr'-\;\; &\;\;\int
 H^I_j(r,r')\delta_{+}(r',r'')dr\cr
  &+\int P^I_J(r,r')H^J_j(r',r'')dr'=0\cr
}\eeqs
and
\beq
        \int H^I_i(r,r')H^{\dag i}_J(r',r'')dr'+\int P^I_K(r,r')P^K_J(r',r'')
 dr' -P^I_J(r,r'')=0
\eeq
The equations of motion and the constraints above define the classical hadron
theory equivalent to spherical QCD in the large $N_c$ limit.

Any static solution to these equations of motion and constraints
 can be used as the starting point for a
semiclassical expansion of spherical QCD. The most obvious solution is the
vacuum:
\beq
        M^i_j(r,r')=0\quad H^I_i(r,r')=0\quad P^I_J(r,r')=0.
\eeq

The small fluctuations around this vacuum will describe the mesons in our
approximation. We are currently interested in the heavy--light mesons described
by the field $H^I_i$. ( It should also be possible to get static
 solutions,solitons, that deviate from the vacuum by a finite amount,
describing
 for
example baryons with one heavy quark).

Expanding around the vacuum, we find that the equation for $H$ decouples from
the others to linear order:
\beq
i\d H^I_i(r,r')/d u
= \half {\tilde\alpha}\int ds' H^I_i(r,s')G(s',r)\delta_{-} (s',r')
+\int  ds'   H^I_i(r,s') h(s',r').
\eeq
The constraint on $H$ becomes to linear order,
\beq
        \int H^I_i(r,r')\delta_{+}(r',r'')dr'=0.
\eeq

Now, $H(r',r)$ is the wavefunction of the heavy--light meson; we should
expect this to be the product of a heavy quark wavefunction concentrated at
the origin, and a light quark wave function.
The ansatz $H^I_i(r',r)=\delta(r')\psi(r)c^I_i$ is therefore reasonable.
The constant $c^I_i$ will determine the internal quantum numbers ( spin,flavor)
that label the degenerate levels of the meson. We get
\beq
i\d \psi(r')/d u
= \half {\tilde\alpha}\int ds' \psi(s')G(0,s')\delta_{-} (s',r')
+\int  ds'   \psi(s') h(s',r').
\eeq
Now recall that $h(s',r')=-h(r',s')$, $\delta_{-}(s',r')=\delta_{+}(r',s')$ so
 that
\beq
-i\d \psi(r')/d u
= -\half {\tilde\alpha}\int ds' \delta_{+} (r',s')\psi(s')G(s',0)+\int  ds'
 h(r',s')\psi(s') .
\eeq

Thus we arrive at the equation for stationary states,
\beq
-\half[i\pdr_r+k^2r^{-1}(i\pdr_r)^{-1}r^{-1}]\psi(r)+ \int
 \delta_{+}(r,s)V(s)\psi(s) ds=E\psi.
\eeq\label{eigen}
Here, $V(r)=-{\tilde \alpha\over 2r}$ is the Coulomb potential and $E$  the
 binding energy of the meson.
 This is to be supplemented by the constraint
\beq
        \int \delta_{-}(r,s)\psi(s) ds=0.
\eeq

 The above pair of integral equations  will
determine the meson wavefunction and masses in our model. They describe
 the propagation of a light quark in the spherically symmetric
potential created by the heavy quark. Since there are no gluon
self--interactions within our approximations, this potential  has turned out to
be the Coulomb potential. Therefore the eigenvalue problem we get is scale
invariant.

In fact we should expect the  scale invariance to be  broken by quantum
effects;
no scale invariant equation can have a discrete spectrum of bound states.
In the next section we will perform a renormalization of the coupling constant
 which will make sure that the ground state energy is finite.

 If we expand
\beq
        \psi(r)=\int_0^{\infty} u_{\lambda}(r) \tilde \psi(\lambda) d\lambda
\eeq
the constraint is automatically satisfied.  The remaining equation can also be
 simplified,
\beq
        \lambda \tilde\psi(\lambda)+\half{\tilde \alpha}\int_{0}^{\infty}
 K(\lambda,\lambda')\tilde\psi(\lambda')d\lambda'=E\tilde\psi(\lambda)
\;\;{\rm for}\;\; \lambda>0.
\eeq
Here
\beq
        K(\lambda,\lambda')=\int dr  V(r) u^*_{\lambda}(r)u_{\lambda'}(r)
\eeq
is the integral kernel of the potential in the basis diagonalizing $h$. This
 is  the analogue of 't Hooft's integral equation.  Using the explicit form of
 the functions $u$ we can simplify these equations further. (For example if
 $k=1$
 they are spherical Bessel functions; for arbitrary $k$, $u(r)$  can  be
 obtained as a power series.) Our result is then  similar to a  Wiener--Hopf
 integral
 equation. The two dimensional analogue of this equation is discussed in
 ref. \[twodim].

{\it \sect. Renormalization and Beta Function}

The eigenvalue equation of the free hamiltonian
\beq
- \half (i\pdr_r+k^2r^{-1}(i\pdr_r)^{-1}r^{-1})u(\lambda r)={\lambda\over 2}
 u(\lambda r)
\eeq
can also written as the differential equation
\beq
        [-\pdr_r(r\pdr_r u)+{k^2\over r} u]=-i\lambda\pdr_r(ru).
\eeq
The solutions are of the form $e^{i\lambda r}$ times a polynomial in ${1\over
 i\lambda r}$. For the special case $k=1$ which we will now study in detail,
\beq
        u(\lambda r)={1\over \surd \pi}[1-{1\over i\lambda r}] e^{i\lambda r}.
\eeq
These functions are orthonormal in the sense that
\beq
        \int_0^{\infty} u^*(\lambda r)u(\lambda'r)dr =\delta(\lambda-\lambda').
\eeq
The  solutions to the constraints are wavefunctions of positive radial
momentum:
\beq
        \psi(r)=\int_0^{\infty} \tpsi(\lambda) u(\lambda r) d\lambda
\eeq
The integral equation is equivalent to minimizing the the energy,
\beq
{\cal E}[\tpsi]={ {\int_0^{\infty} \lambda |\tpsi(\lambda)|^2 d\lambda +
                \int_0^{\infty} V(r)|\psi(r)|^2 dr}\over ||\psi||^2}
\eeq
where $||\psi||^2=\int_0^{\infty}|\tpsi(\lambda)|^2 d\lambda$.

Since it is difficult to solve this problem exactly, we will find the ground
 state energy by a variational principle.

However there is a problem: the ground state energy is divergent.
The potential and kinetic energies are both of dimension $1$ in energy units,
so
 that with $\tpsi_{\mu}(\lambda)=\tpsi(\mu\lambda)$
\beq
        {\cal E}(\tpsi_{\mu})={1\over \mu}{\cal E}(\tpsi)
\eeq
for any $\mu>0$.
If there is one  state with $E<0$ ( an example of which is given below),  there
 are states of arbitrarily
 negative energy. We must introduce a cutoff   and  make the  bare coupling
 coupling constant   depend on  it  such that  the ground state energy remains
 finite as the cutoff  is removed.  If $a>0$ is such a short distance
 cutoff, The regularized energy will be of the form
\beq
        E(a,\talpha_0(a))=a^{-1}E_1(\talpha_0(a))
\eeq
by ordinary dimensional analysis. Define the beta function by
\beq
        \beta({\tilde \alpha}_0(a))=-a{\pdr {\tilde \alpha}_0(a)\over \pdr a}.
\eeq
The condition that $E$ be independent of $a$ is then determines the beta
 function:
\beq
        \beta({\tilde \alpha}_0)=
-{E_1({\tilde \alpha}_0)\over E_1'({\tilde \alpha}_0)}.
\eeq
Such a nonperturbative renormalization scheme was introduced  by Thorn in a
nonrelativistic context\[thorn].

As in field  theory, a naive short distance cutoff is not convenient. It is
 better  to use instead an analogue of dimensional ( or analytic )
 regularization. Define the regularized potential to be
\beq
        V_0(r)=-{{\tilde \alpha}_0\over 2}{\mu_0 \over (\mu_0 r)^{1-\eps}}.
\eeq
We could then compute the regularized energy,
\beq
        E(\mu_0,{\tilde \alpha}_0,\eps)=\mu_0 E_1(\alpha,\eps)
\eeq
from which the beta function can be  determined.

We are not able to calculate the regularized ground state energy exactly;
 instead, we will estimate it by a variational principle. The ansatz
\beq
        \tpsi(\lambda)=\lambda e^{-b\lambda}\;\;{\rm for}\;\; \lambda>0
\eeq
gives
\beq
        E\leq {3\over 2b}-{{\tilde \alpha}_0\over \pi b}\{ {4\over \eps}+4\log
\
   mu_0 b
 -3\}+\;\;{\rm O}(\eps)
\eeq
Minimizing the r.h.s. in the variational parameter $b$  gives
\beq
        \log \mu_0 b=[1-{1\over \eps}+{3\pi\over 8}]+ {3\pi\over 8{\tilde
\alpha
   }_0}
\eeq
so that
\beq
        E_1({\tilde \alpha}_0)\leq -{8{\tilde \alpha}_0\over \pi b({\tilde
\alpha}_0)}
{}.
\eeq
This gives,
\beq
        \beta({\tilde \alpha})\approx -{8\talpha^2\over 8\talpha+3\pi}.
\eeq
This beta function has only one  zero at $\alpha=0$, near which
\beq
{\beta(\talpha)\over \talpha}=-{16\over 3}{\talpha\over 2\pi}+\cdots
\eeq
 so that our renormalized theory is asymptotically free.
This can be compared with the well known one-loop result of perturbation theory
\[narison],\[thooftbeta],\[grosswil],\[politzer]
\beq
        {\beta(\alpha)\over \alpha}= -{11\over 3}{N^2-1\over 2N} {\alpha\over
 \pi}+\cdots
\eeq
Recalling that $\talpha=N\alpha$ we see that ( for large $N$) we have a $16$
 where the one-loop
 result has a $11$; i.e., agreement to about 30 \%.

We found that a more general ansatz such as
\beq
        \tpsi(\lambda)=\lambda(1+c\lambda)e^{-b\lambda}
\eeq
has a minimum at $c=0$; i.e., does not improve the energy. The ansatz
\beq
        \tpsi(\lambda)=\lambda^c e^{-b\lambda}
\eeq
  has a minimum at $c=1.17$ and lowers the energy  by less than a percent.
To get a substantially better estimate of the ground state energy (and hence
the
 beta function) we must solve the problem numerically. It is also of much
 interest to study the excited states. These issues are currently under study
 and we hope to report on them soon.

\vfill\eject
{\it \sect. Appendix: Constraints on Fermion bilinears}

\def\slash#1{ {\gamma\cdot #1}}

\def\for{\;\;{\rm for}\;\;}

In this appendix we will describe how the large $N_c$ limit of fermion
bilinears leads to a classical theory whose phase space is the Grassmannian.
One can think of this as a quantum mechanical analogue of `bosonization'.

Let us consider a set of operators satisfying Canonical anticommutation
relations (CAR):
\beq
        [\chi^{\dag \alpha a},\chi_{\beta b}]_+=
                \delta^{\alpha}_{\beta}\delta^a_b
\eeq
all other pairs of anticommutators being zero. Here $\alpha,\beta=1\cdots N_c$
 while
$a,b$ are all others quantum numbers such as spin, flavor, momentum
etc. It will not matter to us for now what exactly they represent. For
simplicity we will assume that they have a finite range.
A representation for these CAR  can be constructed on the Fermionic Fock space
${\cal F}$ as usual. Now consider the subspace ${\cal F}_0$ of `color singlet'
states:
\beq
        \half [\chi^{\dag\alpha a},\chi_{\beta a}]|s>=0 \for |s>\in {\cal F}_0.
\eeq
(The operators $ \half [\chi^{\dag\alpha a},\chi_{\beta a}]$
generate the algebra $U(N_c)$ of `color'.) In particular this condition implies
that exactly half the one-particle states are occupied.

The color singlet  bilinears
\beq
        \Phi^{a}_{b}= {1\over 2N_c} [\chi^{\dag\alpha a},\chi_{\alpha b}]
\eeq
map ${\cal F}_0$ to itself. In this subspace, they obey the linear constraint
\beq
\Phi^{a}_{a}=0.
\eeq
Also, they satisfy the algebra
\beq
[\Phi^{a}_{b},\Phi^{c}_{d}]={1\over N_c}
[\delta^{c}_{b}\Phi^{a}_{d}-\delta^{a}_{d}
                                                \Phi^{c}_{b}].
\eeq
The meaning of this is that $N_c\Phi^{a}_{b}$ form a representation of
the Lie algebra of the Unitary group on ${\cal F}_0$. This is an irreducible
representation so that the only operators in ${\cal F}_0$ that commute with all
the $\Phi^{a}_{b}$ are multiples of the identity.

 The $\Phi^{a}_{b}$ form a complete set of observables of the  quantum
system whose Hilbert space is ${\cal F}_0$. Within the space of color singlet
states, any observable can be written in terms of the $\Phi^{a}_{b}$.
However, they  are not independent variables; they satisfy a quadratic
constraint between states in ${\cal F}_0$:
\beq
        \Phi^{a}_{c}\Phi^{c}_{b}=
({1\over 4}+{1\over 4N_c}\delta^{c}_{c})\delta^{a}_{b}+
{1\over
2 N_c}\delta^{c}_{c}\Phi^{a}_{b}.
\eeq
To derive these one just has to reorder the factors so that the color singlet
 condition on the states can be used.

In the large $N_c$ limit this becomes just
\beq
        \Phi^{a}_{c}\Phi^{c}_{b}=
{1\over 4}\delta^{a}_{b}.
\eeq

  Now, if we define
the normal ordered bilinear
\beqs{
M^{a}_{b}&=\Phi^{a}_{b}+\half\eps^{a}_{b}\cr
}\eeqs
we will get in the large $N_c$ limit the constraint
\beq
        M^{a}_{c}M^{c}_{b}-
{1\over 2}(M^{a}_{c}\eps^{c}_{b}+
                        \eps^{a}_{c}M^{c}_{\be})=0.
\eeq
To derive the constraints  in the  text we must replace $\chi$ by
 $\pmatrix{\chi\cr Q}$ and the matrix $\Phi$ by $\pmatrix{M-\half\eps&
 H^{\dag}\cr H&P-\half\delta\cr}$.

In the large $N_c$ limit we get as dynamical variable  a hermitian traceless
matrix whose square is a multiple of the identity.
The phase space, which is the set of all such matrices, is the `Grassmannian'.
The commutation relations now are replaced by Poisson brackets on the classical
variables.
We can recover the finite $N_c$  theory by `quantizing'  this classical theory
whose phase space  is the Grassmannian. The wavefunctions of the quantum theory
can be
chosen to be some sort of functions on the Grassmannian. The precise statement
is that the wavefunctions are holomorphic sections  of a line bundle on the
Grassmannian.( The Grassmannian is a complex manifold, so the concept of
holomorphicity makes sense.) This description of the wavefunctions is analogous
to the coherent state picture of the harmonic oscillator. Line bundles on the
Grassmannian are labelled by an integer ( Chern class) which can be identified
with $N_c$. For each $N_c$
therefore we have one quantum theory; the Hilbert space of this quantum theory
can be shown to be just ${\cal F}_0$. This way we can recover the finite $N_c$
theory as a quantization of the classical theory on the Grassmannian. In
particular this construction shows that there are no other constraints on the
$\Phi^{a}_{b}$ in the large $N_c$ limit.

We will encounter some divergences  in extending this to the infinite
dimensional case. However once we deal with normal ordered operators, the
divergences can be handled consistently. A more formal approach is to define
the infinite dimensional Grassmannian with certain  convergence conditions and
to construct the space ${\cal F}_0$ as a space of sections of line bundles on
it. \[pressley]\[mickraj]. We do not need this construction for the purposes of
this paper.

{\it Acknowledgements}

We thank G. Ferretti and Z. Yang for discussions. This work was supported in
part by the US Department of Energy, Grant No. DE-FG02-91ER40685.

{\it References}

\noindent\thooft. G.'t Hooft, Nucl. Phys. B75,461 (1974).

\noindent\witten. E.Witten, Nucl. Phys. B160,57 (1979).

\noindent\N. I.Bars, Phys. Rev. Lett. 36,1521 (1976).

\noindent\pauli. K.Hornbostel, S.Brodsky, H.Pauli, Phys. Rev. D41, 3814 (1990).

\noindent\iw. Isgur N. and Wise M.B., Phys. lett. B232, 113 (1989); B237, 527
(1990); Phys. Rev. Lett., 66, 1130 (1991);Nucl. Phys. B348, 278 (1991).

\noindent\georgi. H.Georgi, Phys.Lett. B240, 447 (1990); Nucl.Phys. B348, 293
(1991).

\noindent\falk. A. Falk, H. Georgi, B. Grinstein, and M. B. Wise, Nucl. Phys.
 B343, 1 (1990).

\noindent\slater. J.C.Slater,{\it  Quantum Theory of Atomic Structure, Vol. 1
and 2},
  McGraw Hill Book Co., Inc. (1960).

\noindent\twodmeson. S.G.Rajeev, `` Two Dimensional Meson Theory '',in 1991
{\it
Summer School in High Energy Physics and Cosmology},edited by E.Gava et al,
World Scientific (1992); P.F.Bedaque, I.Horvath and S.G.Rajeev, Mod. Phys.
Lett.
 A7,35,3347 (1992); S. Guruswamy and S. G. Rajeev, to appear in Mod. Phys.
Lett.
 (1992).

\noindent\brodsky.  S. J. Brodsky and  H. C. Pauli,
{\it Light Cone Quantization of Quantum Chromodynamics}
Invited lectures given at 30th Schladming Winter School in Particle
Physics: Field Theory, Schladming, Austria, SLAC-PUB-5558, Jun 1991.

\noindent\wilson. R. J. Perry, A. Harindranath and K. G. Wilson Phys. Rev.
Lett.   65,2959 (1990).

\noindent\dirac. P. A. M. Dirac, {\it Principles of Quantum Mechanics} Oxford
University Press, Oxford (1953).

\noindent\twodim. B. Grinstein and P. Mende Phys. Rev. Lett. 69,1018 (1992);
M. Burkardt and E. S. Swanson, Phys. Rev. D46, 5083 ( 1992).

\noindent\thorn. C. Thorn, Phys. Rev. D19, 639 (1979); See also  K. Huang, {\it
 Quarks Leptons and Gauge Fields} p. 214, World Scientific, Singapore (1982).

\noindent\narison. S. Narison, Phys. Rep. 84, 263 (1982).

\noindent\thooftbeta. G. 't Hooft, unpublished remark at the Marseille Conf. on
 Yang--Mills Theories (1972).

\noindent\grosswil. D. J. Gross and F.  Wilczek, Phys. Rev. Lett. 30, 1323
 (1973)

\noindent\politzer.  H. D. Politzer, Phys. Rep. 14C, 129 (1974).

\noindent\pressley. A. Pressley and G. Segal, {\it Loop Groups}, Clarendon
Press, Oxford (1986).

\noindent\mickraj. J. Mickelsson and S. G. Rajeev, Comm. Math. Phys. 116,365
(1988).

\bye